\documentclass[epj,nopacs]{svjour}

\usepackage[english]{babel}
\usepackage{amsmath}
\usepackage{graphicx} 
\usepackage{hyperref}
\usepackage{cite}
\usepackage{upgreek}

\begin{document}

\title{Droplet Migration on Conical Fibers} 

\author{
Droplet Migration on Conical Fibers \inst{1} \and 
Carmen L. Lee \inst{1} \and 
Rafael D. Schulman \inst{1} \and
{\'E}lie Rapha{\"e}l \inst{2} \and
Kari Dalnoki-Veress \inst{1}$^{\text{,}}$ \inst{2}$^{\text{,}}$\thanks{\email{dalnoki@mcmaster.ca}}
}

\institute{Department of Physics \& Astronomy, McMaster University, Hamilton, Ontario, L8S 4M1, Canada \and 
UMR CNRS Gulliver 7083, ESPCI Paris, PSL Research University, 75005 Paris, France}

\date{\today}

\abstract{
The spontaneous migration of droplets on conical fibers is studied experimentally by depositing silicone oil droplets onto conical glass fibers. Their motion is recorded using optical microscopy and analysed to extract the relevant geometrical parameters of the system. The speed of the droplet can be predicted as a function of geometry and the fluid properties using a simple theoretical model, which balances viscous dissipation against the surface tension driving force. The experimental data are found to be in good agreement with the model.
}

\maketitle

\section{Introduction}

Spontaneous water transport systems at small lengthscales are a crucial feature for the survival of many living organisms and have been widely studied. In some cases, the aim is to remove excess water. For instance, water strider legs~\cite{wang2015self} and bird feathers~\cite{duprat2012wetting} have water repelling properties. However, many of the organisms employing water transport do so to collect water. Spider webs~\cite{zheng2010directional}, the backs of desert beetles~\cite{parker2001water}, desert moss structures~\cite{pan2016upside} and cacti spines~\cite{ju2012multi,guo2015experimental,luo2015theoretical, malik2015cactus} help these plants and animals collect the water necessary for their survival. In the case of cacti, water collection is based on a simple mechanism: when fog condenses at the tip of a conical cactus spine, a droplet forms that spontaneously migrates towards the widest end of the spine under the action of surface tension as a driving force. Many recent artificial water harvesting systems have been inspired by this mechanism~\cite{li2013structured,ju2013bioinspired,heng2014branched,peng2015magnetically,cao2014facile,xu2016fogcollect, tak2020}, with the intent of fighting drought in arid environments. However, there are still unanswered questions about the droplet dynamics in this system. 

A standard framework to describe droplets on cylindrical fibers has been developed by Caroll~\cite{carroll1976accurate,carroll1986equilibrium}, who also highlighted the existence and the transition between two possible equilibrium configurations for a droplet on a fiber: the \emph{asymmetric clam shell} configuration, in which the droplet is in contact with one side of the fiber, and the\emph{ axisymmetric barrel configuration}, in which the droplet envelopes the fiber. Drop transport along fibers may be induced by several means; previous works have used gradients in coating \cite{bai2010direction}, temperature gradients \cite{hou2013temperature,yarin2002motion} and gravity \cite{chen2013bioinspired,gilet2009microfluidics,gilet2010sliding,kliakhandler2001viscousbeads} to control the motion of droplets on cylindrical fibers. Lorenceau and Qu{\'e}r{\'e}~\cite{lorenceau2004drops} were the first to propose a theoretical model, based on Carroll's framework, to explain barrel shaped droplet dynamics on conical fibers. They showed that a gradient in Laplace pressure along the fiber will drive the motion of the droplet: the thicker the fiber is, the smaller the Laplace pressure will be, thus creating a spontaneous migration of the droplet towards thicker regions of the fiber. Their work is focused on large drop and fiber systems (characteristic length scale of approximately 1 mm), at which gravity must be taken into account and for a small range of relative drop sizes, \textit{i.e.} the drop size compared to the fiber radius. The experimental results from Lorenceau and Qu{\'e}r{\'e} focus on the case where the drop size is comparable to the fiber radius and therefore, the drop is quasi-cylindrical, along with considering the theoretical explanation for when the drop is quasi-spherical. Other recent theoretical, simulation and experimental works on this topic~\cite{michielsen2011gibbs,jian2007directional,liang2015drops,tan2016investigation,shanahan2011behaviour,tak2020,tak2020b} have focused on large length scales, and in some of these studies, investigated where gravity fully balances the surface tension driving force. Smaller length scales have been explored in a more recent study, where the data was analysed with the model presented by Lorenceau and Qu{\'e}r{\'e}~\cite{li2013fastest}. In contrast to the model presented by Lorenceau and Qu{\'e}r{\'e}, others have modelled the driving force being due to surface tension acting on the point of contact between the fiber and droplet \cite{shanahan2011behaviour,tak2020}. Moreover, the migration of clam shell drops has also been investigated~\cite{lv2014substrate,mccarthy2019dynamics}. In addition, the spontaneous migration of drops inside a conical tube or a wedge has been studied recently and theoretical models were proposed to describe this case~\cite{jian2007directional,reyssat2008imbibition,renvoise2009drop,bush2010tweezer,wang2013trapped,reyssat2014drops,ruiz2018statics}. Even if the geometry differs, the same forces are at play: gravity and gradients in the Laplace pressure. More generally, the asymmetry created by cone-like structures has been an inspiration in the field, for instance in elastocapillarity, where a liquid drop between two elastic fibers or thin sheets can make them coalesce or separate~\cite{bico2004adhesion,duprat2011dynamics,aristoff2011elastocapillary,taroni2012multiple,bico2018elastocapillarity}.  

In the present work, we study the migration of highly viscous, totally wetting, and thus barrel-shaped, droplets on conical glass fibers on length scales at which gravity can be neglected. We explore a large range of drop sizes relative to the fiber radius. In this droplet/conical-fiber system, the only forces acting on the droplet are a driving force originating from surface tension and the viscous shear force. We develop a simple theoretical model to predict the droplet speed as a function of the radius, the gradient of the radius, droplet volume, as well as the surface tension and viscosity of the fluid. The model, though different to the that presented by Lorenceau and Qu{\'e}r{\'e}, has similar ingredients.  

\section{Experimental Methods}
\label{sec:experiment}
The glass fibers were prepared by pulling standard borosilicate glass capillary tubes with an outer diameter of 1 mm in a magnetic micropipette puller (Narishige PN-30). The resulting shape of the transformed pipette was a nearly conical fiber with a  changing gradient and tip size of tens of micrometers. We note that though the fibers are ``trumpet'' shaped, with a gradient in diameter that increases weakly as the diameter increases, on the length scale of the droplets they are conical.
 The variations of diameter and gradient were unique to each pipette. Glass was chosen because it presents a well controlled and smooth surface. Three different silicone oils, i.e. poly(dimethyl siloxane) (PDMS), were used: vinyl terminated PDMS, silanol terminated PDMS, and vinyl terminated copolymer (0.3-0.4 $\%$ vinyl{\-}methyl{\-}siloxane)-dimethyl{\-}siloxane (Gelest). The respective kinematic viscosities are 5000 cSt, 2000 cSt and 1000 cSt and a surface tension of $\gamma=$ 22 mN/m. Silicone oils were the most appropriate liquids in this case, as they have well controlled viscosities, are non-volatile, and chemically stable. These oils also totally wet glass which means that droplets have a zero equilibrium contact angle with the fibers. 

\begin{figure}[t]
%\begin{center}
     \includegraphics[width=1\columnwidth]{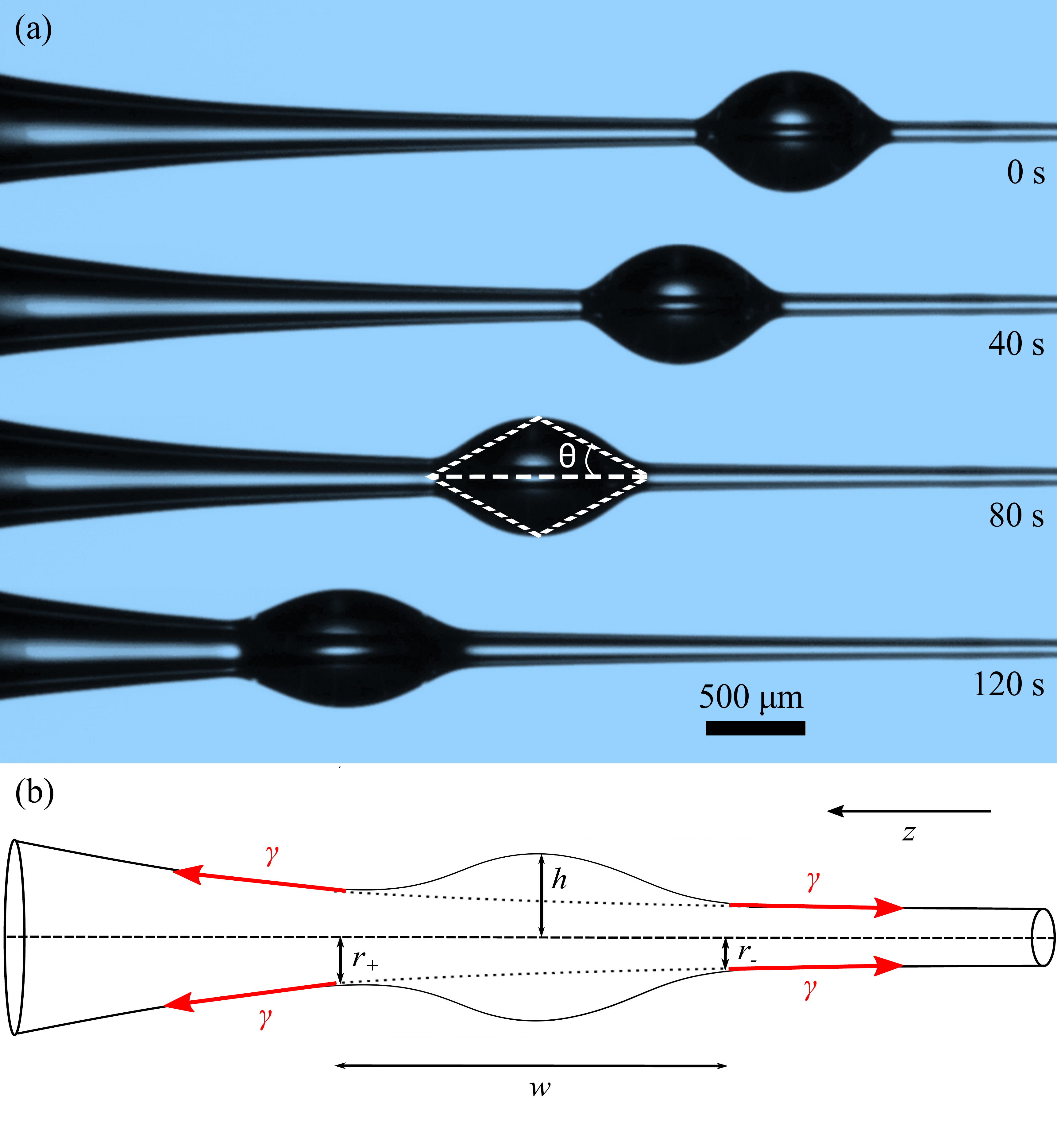}
%\end{center}
\caption{(a) Microscope images of  droplet migrating along a conical fiber at different times. The third panel is annotated with the wedges defined in the model (see section~\ref{sec:model}).  (b) Schematic of a barrel-shaped droplet on a pre-wetted conical fiber with the relevant lengths identified. The surface tension $\gamma$ acting as a driving force on the droplet is also shown.}
\label{fig1}
\end{figure}

A glass fiber is cleaned using acetone and methanol to remove any dust particles. It is then fixed in a horizontal orientation and a small droplet of PDMS is placed close to the tip of the fiber as shown in Fig.~\ref{fig1}(a) at $t=0$ s. The droplet is first produced using another micropipette, which then deposits it on to the fiber by brushing the droplet perpendicularly on the top of the fiber. Once the droplet is deposited, its motion is recorded from above using an optical microscope. Snapshots of the resulting time series of the motion is shown in Fig.~\ref{fig1}(a). An average frame rate of 1 image per second is used. The recording continues until either the droplet exits the field of view or it loses its barrel shape and axial symmetry, which happens when the fiber radius becomes large compared to the droplet size. The first two droplets migrating on an as-cleaned fiber coat the fiber with a thin film of PDMS. Here we focus mainly on subsequent droplets, as we wish to study droplet motion on fibers which are pre-wet by a homogenous thin liquid film. Subsequently, the migration of several droplets of different volumes are recorded and analysed.  

To ensure that gravity had no impact on the measurements, the experimental set-up (including the conical fiber and the microscope) was tilted at an angle and the experiment was reproduced in this configuration. The resulting data presented no difference with the rest of the recorded data, thus gravitational effects are overwhelmed by surface tension driven forces. The dominance of surface tension over gravity in the experiments presented is to be expected since the capillary length of the silicone oils used is $l_{\textrm{c}}=\sqrt{\gamma/(\rho g)} \sim 1.5$~mm; with density $\rho\approx 965$~kg/m$^3$, and $g$ the acceleration due to gravity. 

Several parameters of interest are extracted from the videos: the fiber radius $r$, the radius gradient ${dr}/{dz}$ as a function of the axial coordinate $z$, as well as the droplet properties, height $h$, volume $\Omega$, width $w$, and position at all times. Some of these parameters are denoted in Fig.~\ref{fig1}(b). All parameters are obtained by direct image analysis. The droplet position is retrieved by averaging the $z$-position of the maximum and minimum of parabolas fitted respectively to the top and bottom of the detected edge of the droplet. The speed $v$ is calculated as the numerical time derivative of the droplet position.

\begin{figure}[]
%\begin{center}
     \includegraphics[width=1\columnwidth]{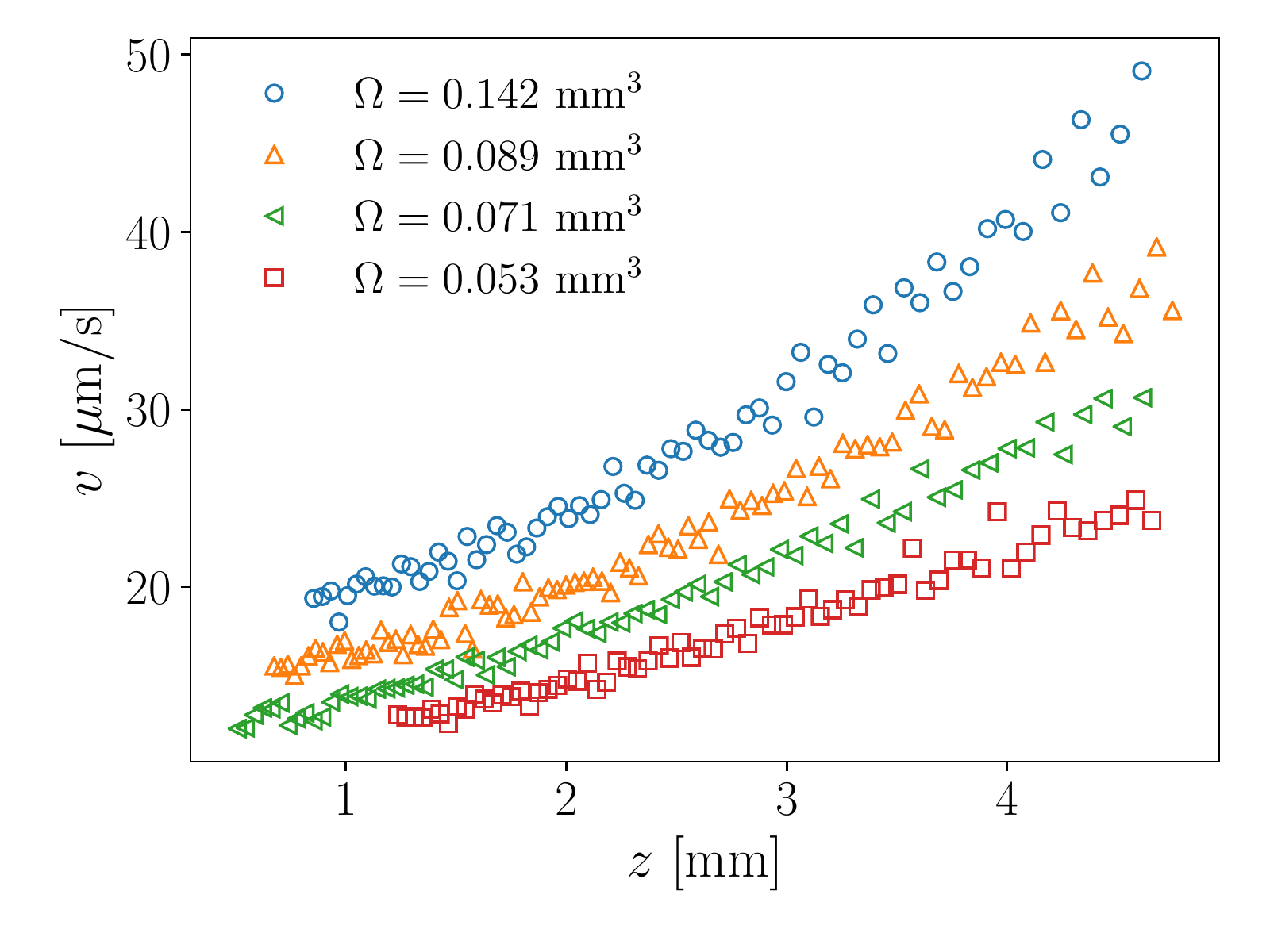}
%\end{center}
\caption{The speed $v$ of several droplets migrating one after another along the same conical fiber plotted as a function of their position on the fiber $z$. The different markers represent droplets of different volumes $\Omega$.
}
\label{fig2}
\end{figure}

As the goal of this study is to find a comprehensive expression for $v$ as a function of all the other variables, a first step is to look  at how the speed varies with the other parameters of the system. A plot of $v$ as a function of the droplet position (Fig.~\ref{fig2}) shows the speed of four droplets of different volumes that migrated on the same pipette. From this plot, it is evident that the speed increases with position. However, we also know that both the radius and the gradient of the fiber increase with  position. Therefore, the raw data does not allow us to dissociate the effect of each parameter. A second observation is that the speed increases with the droplet volume. Since there are several variables which influence the droplet speed, it is necessary to develop a model in order to attain a comprehensive expression for $v$ as a function of all the relevant parameters of the system.

\section{Theoretical Model}
\label{sec:model}

In order to develop a model, we consider all the forces applied on the droplet. Gravity is negligible, verified both by the experiment (see Section~\ref{sec:experiment}), and the fact that the length scale of the experiments are well below the capillary length. We assume that inertial forces are negligible, as the Reynolds number is on the order of $10^{-6}$. Two main contributions remain: the driving force caused by the surface tension $\gamma$  of the silicone oil, as depicted in Fig.~\ref{fig1}(b), and the viscous dissipation in the liquid. 
The first contribution, which we denote $F_{\gamma}$, is the net surface tension force exerted by liquid-air interface of the pre-wetted fiber onto the droplet volume. It is composed of a difference between two surface tension forces: one acting at the extremity of the advancing side where $r=r_{+}$,  and another acting at the receding side where $r=r_{-}$. We note that here we consider the external forces acting on the droplet (surface tension at the contact lines), rather than the method of internal Laplace pressure gradients as implemented by Lorenceau and Qu{\'e}r{\'e}~\cite{lorenceau2004drops}. A simple approximation gives the following final expression for $F_{\gamma}$: 
\begin{equation}
F_{\gamma} \sim 2 \pi \gamma r_{+} - 2 \pi \gamma r_{-}  \sim \gamma w \frac{dr}{dz}.
\label{gamma_eq}
\end{equation}
We note that this simple approximation ignores the difference in the advancing and receding contact angles~\cite{tak2020}. The second contribution, denoted $F_{\eta}$, results from the viscous force at the solid-liquid interface. In order to attain a simple expression for this quantity, we approximate the droplet shape as two joined wedges drawn in the third panel of Fig.~\ref{fig1}(a). Although a crude approximation, it should suffice for quantifying the dissipation at the level of scaling. In this case, the viscous force can be evaluated by integrating the shear force over the entire liquid-solid area ($A_\mathrm{ls}$) beneath the wedge~\cite{de2003capillarity}: 
\begin{equation}
F_{\eta} \sim \iint_{A_\mathrm{ls}} \eta \frac{dv}{dy} \biggr\rvert_{y=0}  \sim  \frac{\eta r v}{\tan(\theta) } \, ,
\label{eta1_eq}
\end{equation}
where $y$ represents the radial coordinate, which is equal to 0 when the point of interest is at the center of the fiber, $\eta$ the dynamic viscosity, and $\theta$ is angle of the wedges which approximate the droplet. Here, we have neglected any prefactors and the logarithmic term which truncates the integral in viscous wedge dissipation (see Ref.~\cite{de2003capillarity}).
The only parameter from this expression that has not directly been measured in the experiment is $\tan(\theta)$. Using simple trigonometry, we have:
\begin{equation}
\tan(\theta)=\frac{h}{w/2} \, .
\label{theta_eq}
\end{equation}
Substituting Eq.~\ref{theta_eq} back into Eq.~\ref{eta1_eq}, a final expression for $F_{\eta}$ is obtained: 
\begin{equation}
F_{\eta} \sim  \frac{\eta r v w}{h} \, .
\label{eta2_eq}
\end{equation}
In the absence of gravity and inertia, we must have $F_\eta \sim F_\gamma$. Thus, equating Eq.~\ref{gamma_eq} and Eq.~\ref{eta2_eq} yields an expression for the speed as a function of all the other relevant parameters: 
\begin{equation}
v \sim \frac{\gamma}{\eta} \frac{h}{r} \frac{dr}{dz} \, .
\label{eqforce2_eq}
\end{equation}

\begin{figure}[]
%\begin{center}
     \includegraphics[width=1\columnwidth]{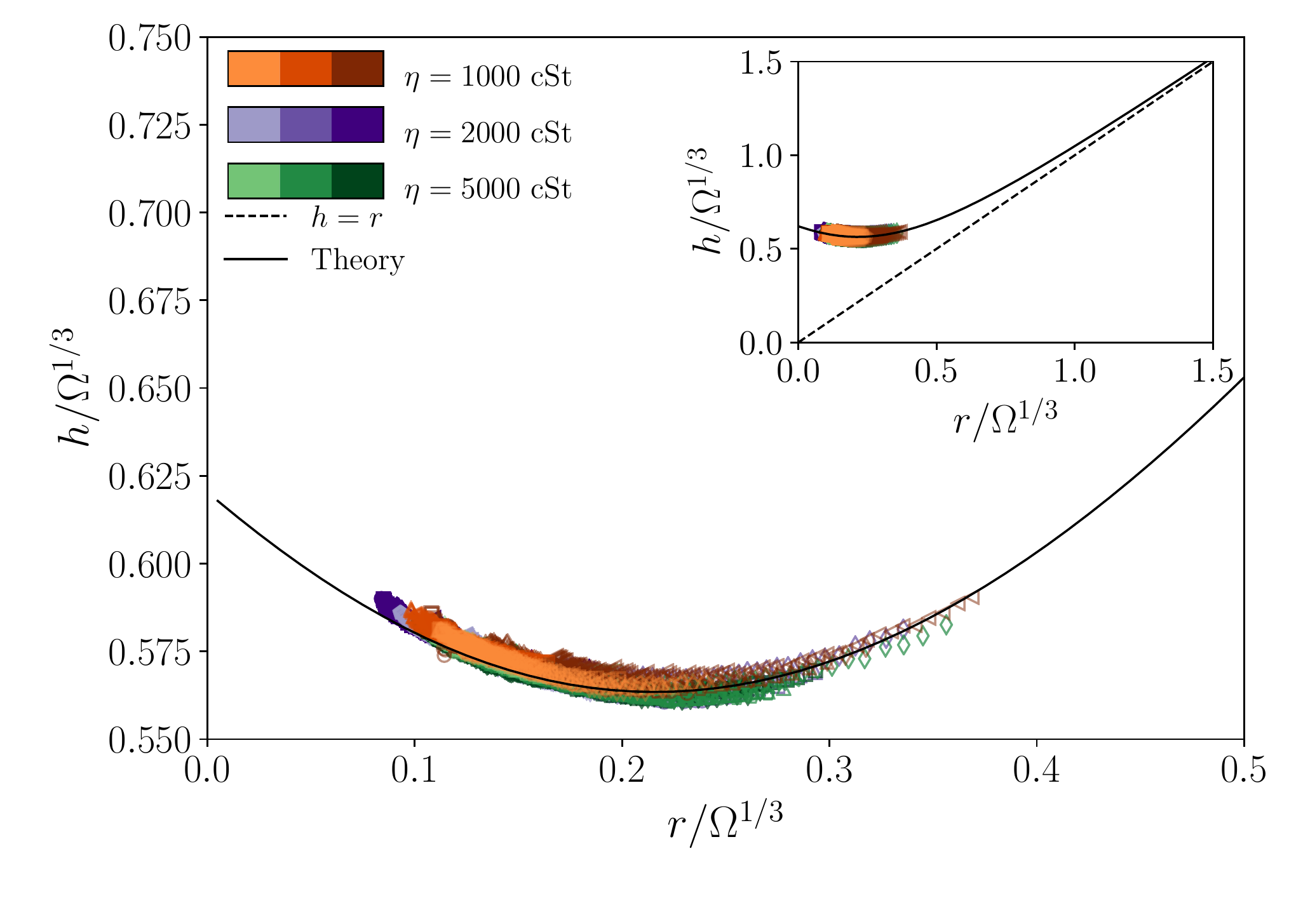}
%\end{center}
\caption{The reduced height $h/\Omega ^{1/3}$ of all studied droplets plotted as a function of the reduced radii of the fibers $r/\Omega ^{1/3}$, on which the droplets are migrating. The markers show the experimental data for all droplets. The different colors for the markers represent viscosities: 1000 cSt (orange), 2000 cSt (purple) and 5000 cSt (green); different shades of a colour correspond to different fibers. The solid black line represents the theoretical relationship between $h/\Omega ^{1/3}$ and  $r/\Omega ^{1/3}$, as derived by Carroll~\cite{carroll1976accurate,carroll1986equilibrium} with no fitting parameters. The inset in the image shows the experimental data, the theoretical relationship, and the dashed black line of equation $h=r$ shows that the predicted $h$ tends towards $r$ for large fiber radii on a larger scale compared to the main figure.
}
\label{fig3}
\end{figure}

The droplet height can in principle be obtained from an exact equation of the shape of a droplet on a fiber, assuming the quasi-static approximation. For a barrel droplet on a cylindrical fiber it has been shown that there is a non-trivial dependence on the fiber radius~\cite{roe1975wetting,carroll1976accurate,carroll1986equilibrium,mchale2001shape}. The relationship between $h$ and $r$, as derived by Carroll~\cite{carroll1976accurate, carroll1986equilibrium}  is shown as a solid black line in Fig.~\ref{fig3} with reduced coordinates $h/\Omega ^{1/3}$ and $r/\Omega ^{1/3}$ and is non-monotonic. The dashed black line shown in the inset of Fig.~\ref{fig3} represents the asymptotic regime in which $h=r$. For large $r$, $h$ tends towards $r$ as the droplet flattens and takes the shape of a cylinder enveloping the fiber. For small $r$, $h$ is close to that of a droplet in a quasi-spherical regime, where $h /\Omega^{1/3} \approx 0.62$ as $r \rightarrow 0$ (a droplet can be approximated as a sphere of radius $h$).
Experimentally the parameters in Eq.~\ref{eqforce2_eq} can all be determined easily. $\eta$ and $\gamma$ are known fluid properties and $r$ and $\frac{dr}{dz}$ are geometrical properties of the fiber which can be determined from image analysis as discussed above  (see Section~\ref{sec:experiment}). Lastly, since the droplet height  varies non-trivially with fiber radius, we determine $h$ experimentally at every image frame in the sequence. 

\section{Results and Discussions}

We first turn to the height of the droplet on a fiber. Since the droplet migrates, if we make the quasi-static assumption (i.e. neglect fiber gradient and assume $v=0$), the experiment is equivalent to extracting the height of a droplet as a function of fiber radius. The experimental data is shown in Fig.~\ref{fig3}, with markers corresponding to the three different viscosities: 1000 cSt (orange), 2000 cSt (purple) and 5000 cSt (green); different shades correspond to different conical fibers. To the best of our knowledge this relationship has not been experimentally verified and the agreement with the classic theory is excellent with no fitting parameters. Although for small fiber radii $h$ is close to that of a droplet in a quasi-spherical regime, we do not make the common approximation of $h \sim \Omega^{1/3}$ in this study.

\begin{figure}[b]
%\begin{center}
     \includegraphics[width=1\columnwidth]{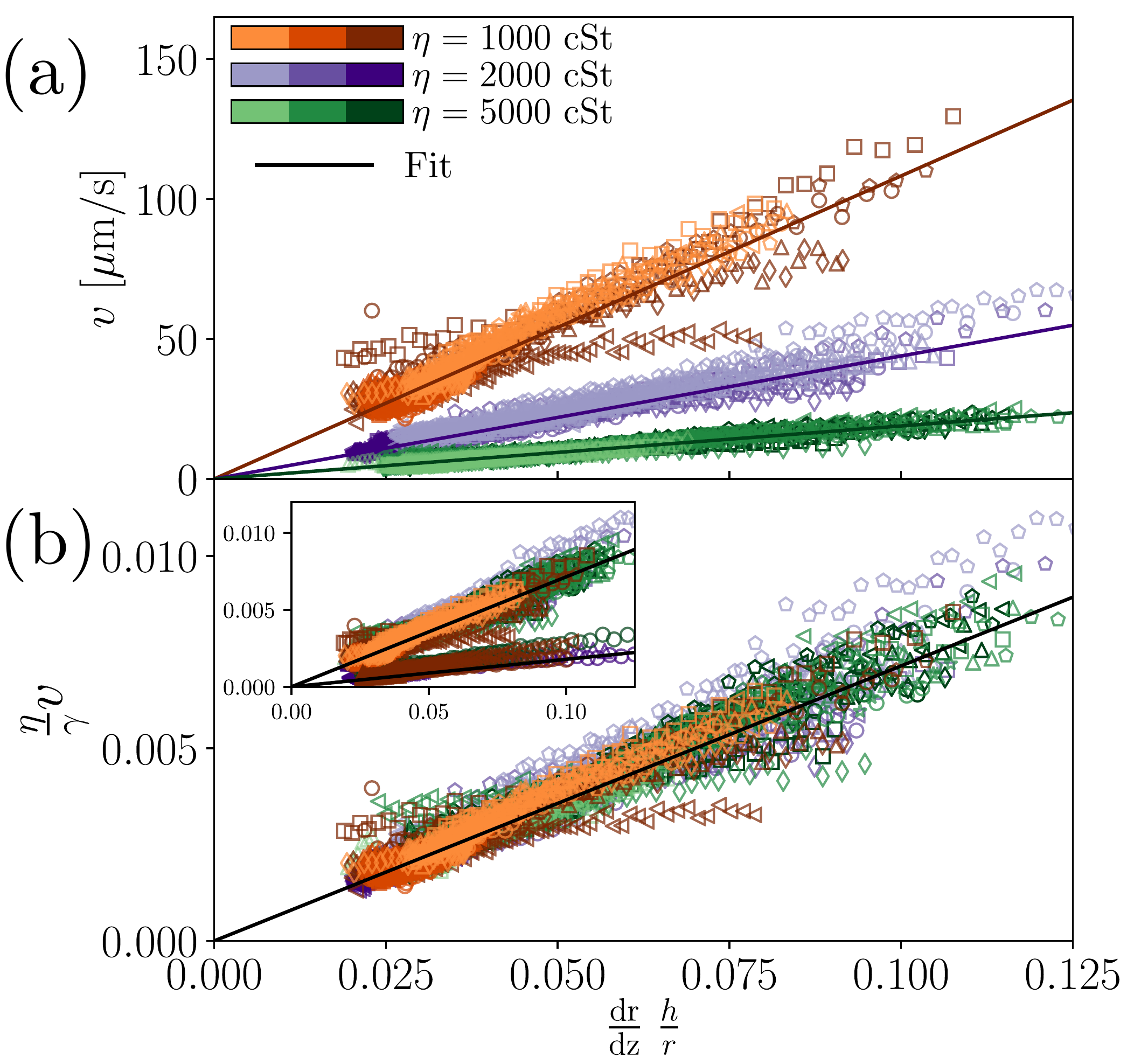}
%\end{center}
\caption{(a) Speed for all studied droplets as a function of the gradient of their respective fibers multiplied by their height and divided by their radii, $ \frac{dr}{dz}\frac{h}{r}$. The different color families represent the viscosity of the PDMS: 1000 cSt (orange markers), 2000 cSt (purple markers) and 5000 cSt (green markers). The different colour shades denote different fibers (3 different fibers for each viscosity). The different types of markers (circle, square, etc) denote different droplets. The lines represent the best fit of the model (Eq.~\ref{eqforce2_eq}) for each viscosity. The best fit lines have slopes of $a \gamma/\eta$, where $a$ was found to be $\sim$0.074$\pm$0.002. (b) Speed normalized by the capillary velocity for all studied droplets as a function of $ \frac{dr}{dz}\frac{h}{r}$. The colour scheme is the same as in Fig 4(a). The straight line represents the best fit of the model to all data with the best fit slope of $\sim$0.074$\pm$0.002. The inset includes the data from the main panel and the speed normalized by the capillary velocity for droplets on a dry fiber as a function of $ \frac{dr}{dz}\frac{h}{r}$. The best fit slope for the data on a dry cone is $\sim$0.018$\pm$0.003. 
}
\label{fig4}
\end{figure}

The droplet speed as a function of experimentally relevant parameters is given by the model prediction presented in Eq.~\ref{eqforce2_eq}.
In Fig.~\ref{fig4}(a) we plot the speed, $v$ as a function of $\frac{h}{r} \frac{dr}{dz}$. The model predicts a straight line through the origin with a slope proportional to  $\frac{\gamma}{\eta}$. Indeed, for all the experiments (52 in total). The experimental data is in excellent agreement with the simple model despite the approximations made (see \cite{tak2020} for a more rigorous theory and simulation). The droplets follow three different trends corresponding to the three different viscosities: 1000 cSt (orange markers), 2000 cSt (purple markers) and 5000 cSt (green markers). For each viscosity, three different fibers of varying radii and gradients were used to diversify the experimental conditions. They correspond to the different colors shades of the markers. On each single fiber, several droplets migrations (6 on average) were recorded once the fiber was pre-wet. In Fig.~\ref{fig4}(a) each different type of marker (circle, square, etc.) corresponds to a different droplet. According to Eq.~\ref{eqforce2_eq}, these straight lines have slopes that are inversely proportional to the respective viscosities of the fluids but otherwise have the same prefactor. The prefactor itself is the result of the multiplication of the surface tension $\gamma$ (the same for the three oils that were used) and a numerical coefficient of order of $10^{-1}$. 

The comparison between model and data can be tested further by dividing the speed by the capillary velocity of the oil $\gamma/\eta$, which normalizes the data by the corresponding viscosities. The expectation is then to obtain a single trend for all droplets, in this case a straight line going through the origin with a slope equal to the prefactor discussed earlier: a numerical prefactor which we find to be equal to 0.074$\pm$0.002 by fitting the data. We see in Fig.~\ref{fig4}(b) that all data collapse in agreement with Eq.~\ref{eqforce2_eq}. The agreement between the data and the model further demonstrates the robustness of the model.  

Limitations of the model approximations could be observed in the experiments. For instance, when the droplet continues to migrate towards the larger end of the fiber, there is a point where the droplet becomes asymmetric which results in a deviation of the data from the expected straight line when plotted as in Fig.~\ref{fig4}(b). Failure of the model can also be observed for droplets with extreme volumes when the volume is sufficiently large and gravity cannot be neglected. This deviation happens when the characteristic dimension of the droplet is comparable to the capillary length of the liquid.

A further interesting observation can be made about the first droplets to migrate on each fiber. In that case the droplets are moving on a dry fiber. The resulting data for these droplets are still found to collapse onto a straight line going through the origin when plotted as in the inset in Fig.~\ref{fig4}(b). The only difference is that the slope of this line is smaller from the slope of the wet fiber data by a factor of approximately 4. The altered dynamics is reasonable because in the absence of a pre-wetting film there is greater viscous dissipation at the advancing contact line and furthermore one can expect a modification to the driving force depicted in Fig.~\ref{fig1}(b). Regardless, the deviation results in a constant pre-factor so that the overall scaling remains unchanged.  

The work presented here differs from the experiments presented by  Lorenceau and Qu{\'e}r{\'e} in 2004~\cite{lorenceau2004drops} and the work by Li and Thoroddsen in 2013~\cite{li2013fastest} in several ways. First, in our experiments, gravity is negligible because we consider droplets with characteristic sizes smaller than the capillary length, while Lorenceau and Qu{\'e}r{\'e} studied larger droplets. Second, here we focus on a broad range of relative droplet sizes in the quasi-spherical regime. In contrast, Lorenceau and Qu{\'e}r{\'e} focus on experimental results in the quasi-cylindrical regime. Third, our theoretical model differs from the model built by Lorenceau and Qu{\'e}r{\'e} and used by Li and Thoroddsen. While we also predict the droplet speed using a balance of both the viscous dissipation and the driving force, we use a different approximation for the dissipation and define our driving force through the tensions at the contact line. 
%We note, that in the approximation for dissipation we have only considered the viscous shear at the liquid-solid contact line; however, we have also considered dissipation due to the meniscus using models based on Tanner's Law. However, for the system we have considered, the data proves to follow a scaling predicted by a bulk dissipation model rather than a meniscus dissipation model \cite{keiser2017dropfriction}. 
Lastly, in previous works by Lorenceau and Qu{\'e}r{\'e} as well as Li and Thoroddsen, experimental data is provided as evidence for the models. Just as for the model presented here, the authors predict a direct proportionality between the speed as a function of fiber gradient, with a speed that vanishes as the gradient tends to zero. In both studies~\cite{lorenceau2004drops,li2013fastest} when comparing data to the theory, the data is consistent with a linear dependence of speed on the fiber gradient, but inconsistent with the prediction that the speed must vanish when the gradient vanishes (i.e. the best fit lines do not go through the origin and predict a non-zero velocity when there is no gradient). We have resolved this inconsistency with our model and data.

\section{Conclusions}

In this work, the spontaneous migration of a droplet on a fiber with a radius gradient has been characterized. If gravity and inertia are negligible the speed of the droplet can be predicted as a function of the other parameters of the system. These parameters are geometrical (radius and gradient of the fiber, droplet height) and fluid characteristics (viscosity, surface tension). The predicted speed of droplet migration is based on a simple theoretical model in which the viscous shear force on the droplet balances the surface tension driving force. We find a good agreement between this model and the experiments that were performed using various fiber shapes and droplet volumes. We further validate the non-monotonic dependence of droplet height on fiber radius. One could imagine using the model  as a way to improve future fog harvesting devices inspired by this spontaneous droplet migration mechanism often seen in biological systems.

 \begin{acknowledgement} The authors are grateful to Michael Brook and Dan Chen for providing the silicone oils, and to Tak Shing Chan and Andreas Carlson for extensive discussions. The financial support by the Natural Science and Engineering Research Council of Canada is gratefully acknowledged.
 \end{acknowledgement}

\section*{Author contribution statement}
CF, RDS and KDV designed the research project, CF performed all experiments, all authors contributed to analyzing the data and developing the models, CF wrote the first draft of the manuscript. All authors edited the manuscript to generate a final version and contributed to discussions.

\bibliography{biblio}

\end{document}